\documentclass[final,aps,prl,amsmath,amssymb,superscriptaddress,twocolumn]{revtex4-2}

\usepackage{times}
\usepackage[T1]{fontenc}
\usepackage{blindtext}
\usepackage{amsfonts}
\usepackage{amsmath}
\usepackage{microtype}
\usepackage[usenames,dvipsnames]{xcolor}
\definecolor{shadecolor}{gray}{0.9}
\usepackage{graphicx}
\usepackage{overpic}
\usepackage{booktabs}
\usepackage{tabularx}
\usepackage[ruled,lined]{algorithm2e}
\usepackage{algpseudocode}
\usepackage{lipsum}
\usepackage{natbib}
\usepackage[colorlinks,linktoc=all]{hyperref}
\usepackage[all]{hypcap}
\definecolor{darkblue}{rgb}{0.0, 0.0, 0.55}
\definecolor{darkmidnightblue}{rgb}{0.0, 0.2, 0.4}
\definecolor{dukeblue}{rgb}{0.0, 0.0, 0.61}
\definecolor{zaffre}{rgb}{0.0, 0.08, 0.66}
\hypersetup{citecolor=zaffre}
\hypersetup{linkcolor=zaffre}
\hypersetup{urlcolor=zaffre}
\usepackage[nameinlink,capitalize]{cleveref}
\usepackage{todonotes}

\newcommand{\be}{\begin{equation}}
\newcommand{\ee}{\end{equation}}
\newcommand{\bea}{\begin{eqnarray}}
\newcommand{\eea}{\end{eqnarray}}

\newcommand{\f}[2]{\frac{#1}{#2}}

\DeclareMathOperator*{\argmin}{arg\,min}
\usepackage{printlen}

\definecolor{brotcolor}{rgb}{0.0, 0.5, 0.0}
\definecolor{PSGDcolor}{rgb}{1.0, 0.49, 0.0}
\definecolor{otcolor}{rgb}{0.0, 0.5, 1.0}

\newcommand{\Nodes}{V}

\newcommand{\Nnode}{{|V|}}

\newcommand{\NetMass}{S}
\newcommand{\Inc}{B}

\newcommand{\Inode}{v}
\newcommand{\Inodetwo}{u}

\newcommand{\Iedge}{e}
\newcommand{\Iedgetwo}{{e'}}

\newcommand{\Res}{r}

\newcommand{\Lap}{L}

\crefrangeformat{equation}{eqs. #3(#1)#4-#5(#2)#6}
\crefformat{equation}{Eq.~#2(#1)#3}
\Crefname{equation}{Equation}{Equations} 
\crefname{eqnarray}{eq.}{eqs.}
\Crefname{eqnarray}{Eq.}{Eq.}
\crefname{figure}{fig.}{figs.}
\Crefname{figure}{Fig.}{Figs.}

\usepackage[caption=false]{subfig}

\usepackage[normalem]{ulem} 

\newcommand{\model}{BROT}

\makeatletter
\def\maketitle{
\@author@finish
\title@column\titleblock@produce
\suppressfloats[t]}
\makeatother

\begin{document}

\title{Bilevel Optimization for Traffic Mitigation in Optimal Transport Networks}
\author{Alessandro Lonardi}
\email{alessandro.lonardi@tuebingen.mpg.de}
\affiliation{Max Planck Institute for Intelligent Systems, Cyber Valley, T{\"u}bingen 72076, Germany}
\author{Caterina De Bacco}
\email{caterina.debacco@tuebingen.mpg.de}
\affiliation{Max Planck Institute for Intelligent Systems, Cyber Valley, T{\"u}bingen 72076, Germany}

\begin{abstract}
Global infrastructure robustness and local transport efficiency are critical requirements for transportation networks.  However, since passengers often travel greedily to maximize their own benefit and trigger traffic jams, overall transportation performance can be heavily disrupted. We develop adaptation rules that leverage Optimal Transport theory to effectively route passengers along their shortest paths while also strategically tuning edge weights to optimize traffic. As a result, we enforce both global and local optimality of transport. We prove the efficacy of our approach on synthetic networks and on real data. Our findings on the International European highways suggest that thoughtfully devised routing schemes might help to lower car-produced carbon emissions.
\end{abstract}


\maketitle
\emph{Introduction.}---Transport networks are ubiquitous in nature and engineering, spanning from living organisms to cities and telecommunications.  Many of these systems can be modeled by adaptation rules that follow the principle of minimum energy, regulating edge flows to optimize transportation costs. Examples in biology are plants, whose profiles emerge from a trade-off between minimization of hydraulic resistance and carbon cost \cite{kocillari2021widened}, and leaves, shaped by the interplay of nutrients' transport efficiency and robustness to damage \cite{katifori2010damage, ronellenfitsch2016global, ronellenfitsch2019phenotypes}. 

Similarly, adaptation rules have been employed to model traffic flows in urban transportation by jointly minimizing the energy dissipated by the passengers and the construction cost of the infrastructure \cite{Tero439,lonardi2021designing,lonardi2021multicommodity,lonardi2021infrastructure,adinoyi2021optimal,adinoyi2022sustainable,bonifaci2021physarum}. While these models set forth a first approach to simulate traffic flows using adaptation, they crucially neglect that passengers in a transportation network do not move cohesively to minimize a unique cost. Instead, they choose their routes greedily to maximize their benefit (Wardrop's first principle) \cite{youn2008price, gastner2006optimal, selten2007commuters}.  As a consequence, transport networks may be globally inefficient.

In this work,  we propose a set of adaptation equations to find traffic flows that mitigate congestion, considered as a proxy for global efficiency, while trading off against the shortest routes.

We frame the problem in a bilevel optimization setup, which poses a competition between greedy passengers and a network manager. The passengers minimize their origin-destination path cost seeking for the User Equilibrium \cite{beckmann1956studies} (lower-level problem),  whereas the network manager guarantees global efficiency by mitigating traffic bottlenecks on edges to achieve the System Optimum (upper-level problem), while implicitly accounting for passengers' shortest path.  We tackle the optimization problem by alternating Optimal Transport (OT)-inspired adaptation rules for the lower-level optimization,  and a Projected Stochastic Gradient Descent (PSGD) scheme for the upper-level optimization.

In detail, greedy passenger flows are found by solving a dynamical system that governs the evolution of edge capacities, variables that control passengers allocation, so that these travel on their shortest paths.  Adaptation rules are a well-established mechanism for route assignment on networks \cite{Tero439,lonardi2021designing,lonardi2021multicommodity, bonifaci2012physarum,bonifaci2013short,lonardi2021infrastructure,tero2007mathematical,folz2023noise,hu2013adaptation,ronellenfitsch2016global,kirkegaard2020optimal,adinoyi2021optimal,bonifaci2021physarum,adinoyi2022sustainable} and in continuous domains \cite{baptista2020network,facca2016towards,facca2019numerics,facca2020branching,leite2022revealing}.  Classically, User Equilibrium greedy flows can be found with the Frank-Wolfe algorithm \cite{leblanc1975efficient}, or alternatively with recent methods accounting for passengers' travel budgeting \cite{bao2015travel}. Here, we propose a model that exploits OT theory to prove that, at convergence, passengers move along the shortest path.  Particularly, our dynamical system admits a Lyapunov functional \cite{facca2019numerics} which asymptotically converges to the shortest path (Wasserstein) distance between entry and exit distributions of passengers \cite{bonifaci2012physarum, bonifaci2013short, lonardi2023immiscible}. 

Traffic mitigation is performed by minimizing a quadratic loss function that penalizes edges whose traffic exceeds a prefixed threshold. The minimization problem can be treated analytically by assuming that the network edges are endowed with capacities and weights (resistances) and their flows are the gradient of a scalar potential, as for electrical networks. We derive closed-form gradients for the weights, which can be interpreted as the cost that passengers pay for traveling. In practice, network managers would  implement these weights by strategically designing incentives or disincentives, e.g., assigning road tolls, to encourage passengers to relocate from jammed edges. The task of traffic mitigation has been addressed using several methods. These include belief propagation \cite{po2021futility,xu2021scalable,yeung2019coordinating},  adaptive dynamical networks \cite{jiang2019optimizing},  MCMC schemes \cite{colizza2004network},  cellar automata \cite{tak2019benefit, tak2022adaptive}, and heuristic routing models \cite{gang2006efficient}. 

A bilevel optimization problem similar to the one studied here was solved using message-passing \cite{li2022bilevel}.  While the problem's setting is similar to ours, the methodologies differ since we alternate adaptation rules for the capacities with global descent for the weights, whereas message-passing uses local updates for flows. Our approach outputs individual passengers' optimal paths, whereas the formulation in \citet{li2022bilevel} can only extract aggregate routes.

We find that our method effectively trades off traffic mitigation against the shortest passengers' routes. Namely,  both on synthetic topologies and real roads, it returns optimal transport networks where congestion is heavily reduced. We argue that this result is beneficial for reducing the carbon footprint of roads.  We also show that the uncoordinated actions of network manager and passengers can be counterproductive, i.e., they may increase traffic, with an outcome opposite to that intended.

\emph{Problem.}---We take a network $G(V,E)$ where $M \geq 1$ groups of greedy passengers $i$ can travel from origin nodes $O^i$ (one node per group), to possibly multiple destination nodes $D^i$.  Stationary numbers of entry and exit passengers are stored in a mass matrix with entries $\tilde{S}_v^i > 0$ for each $v = O^i$,  $\tilde{S}_v^i < 0$ for $v \in D^i$, and $\tilde{S}_v^i = 0$ otherwise.  We assume that the system is isolated, i.e., that passengers entering the network must also exit.  This condition is $\sum_v \tilde{S}_v^i =  0$ for all $i$.  When traveling along an edge, passengers pay a cost $\tilde{w}_e > 0$, and lastly, each edge is equipped with a capacity that controls the rate at which passengers $i$ are allocated along each edge $e$---$\tilde{c}^i_e \geq 0$. Intuitively, one could think of capacities as the space occupied by passengers of type $i$, i.e., larger space accommodates more passengers. All problem variables have been introduced with units, however, these can be nondimensionalized to derive scale-independent adaptation rules \cite{si}. We denote dimensional quantities with a tilde, and dimensionless ones without.

\emph{Lower-level optimization.}---The lower-level problem allows us to find the cheapest routes from $O^i$ to $D^i$.  In order to model traffic flows, we introduce the fluxes $F_e^i$, specifying the displacement of $S^i$ along an edge $e$. In analogy with electrical networks, we assume that there exists an auxiliary pressure potential $p_v^i$ on each node $v$ due to index $i$. We interpret them as the travel demand from passengers traveling from $v$. With this, we define the potential-based fluxes for all $e =(u,v)$ and $i$, i.e., Poiseuille’s Law, as 
\begin{equation}
\label{eqn:flux_def}
F_e^i = \frac{c_e^i}{w_e} (p_u^i - p_v^i) \,.
\end{equation}
Fluxes must obey Kirchhoff's law. We can write it as $\sum_e B_{ve} F_e^i = S_v^i$, where $B$ is a conventionally oriented incidence matrix of the network. Substituting \Cref{eqn:flux_def} in Kirchhoff's law, the potential becomes a function of $c$ and $w$, namely $p_v^i = \sum_u (L^{i^\dagger})_{vu} S_u^i$, where $\dagger$ denotes the Moore-Penrose inverse and $L^{i}_{uv} = \sum_{e} (c_e^i / w_e) B_{ue} B_{ve}$ are entries of the network weighted Laplacian.  With this substitution, $F \equiv F(c,w)$ is also a function of only $c$ and $w$, the only independent problem's variables.

For any fixed set of weights,  we write the lower-level problem as
\begin{align}
\label{eqn:j_cost_def}
J(c,w) &= \sum_{ei} w_e |F_e^i| \\
\label{eqn:j_cost}
&\min_{c \geq 0} J(c,w) \,.
\end{align}
The convex OT cost $J$ in \Cref{eqn:j_cost_def} is the sum over $M$ indexes of the $w$-shortest path costs $J^i = \sum_e w_e |F_e^i|$ \cite{bonifaci2012physarum, bonifaci2013short}. Its only minimizer is the overlap of $M$ shortest paths from all $O^i$ to $D^i$, that are found with $c$ using \Cref{eqn:flux_def} and Kirchhoff's law.

\emph{Upper-level optimization.}---The upper-level problem formalizes the task of the network manager of tuning $w$ to mitigate traffic jams triggered by the passengers. We measure traffic by penalizing congested links where $\sum_i |F_e^i|$ exceeds a threshold $\theta \geq 0$, above which infrastructural failures may occur.  Conveniently, we introduce $\Delta_e = \sum_i |F_e^i| - \theta$. 

Analogously to \Crefrange{eqn:j_cost_def}{eqn:j_cost}, for any set of capacities, the upper-level optimization is
\begin{align}
\label{eqn:omega_cost_def}
\Omega(c,w) &=
\frac{1}{2} \sum_e \Delta_e^2 \,H(\Delta_e)\\
\label{eqn:omega_cost}
&\min_{ w \geq \epsilon } \Omega(c,w) \,,
\end{align}
where $H$ is the Heaviside step function. In \Cref{eqn:omega_cost_def}, other objective functions, e.g., the hinge loss can be utilized \cite{li2022bilevel, rocks2019limits}, we do not explore this here.  Furthermore, the weights are constrained to be larger than a small $\epsilon > 0$. This means that passengers cannot profit ($w < 0$), or travel for free ($w = 0$). Practically, this ensures that the Laplacian $L$ is well-defined.

\begin{figure}[t]
\centering
\includegraphics[width=\linewidth]{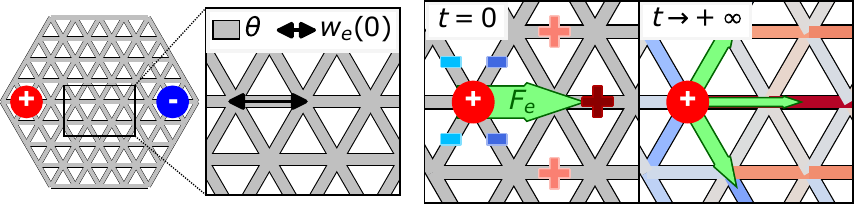}
\caption{Bilevel optimization scheme on a lattice. Entry and exit inflows are the red and blue nodes, respectively.  Initially,  (green) fluxes distribute minimizing the travel cost $w_e(t=0) = \ell_e$, being the length of an edge. If they exceed $\theta$ they get penalized, hence, the network manager tunes the weights to encourage rerouting over more expensive (red), or cheaper (blue) edges (for a companion Fig. \cite{si}).}
\label{fig:example_fig}
\end{figure}

\begin{figure*}[htpb]
\centering
\includegraphics[width=1.0\linewidth]{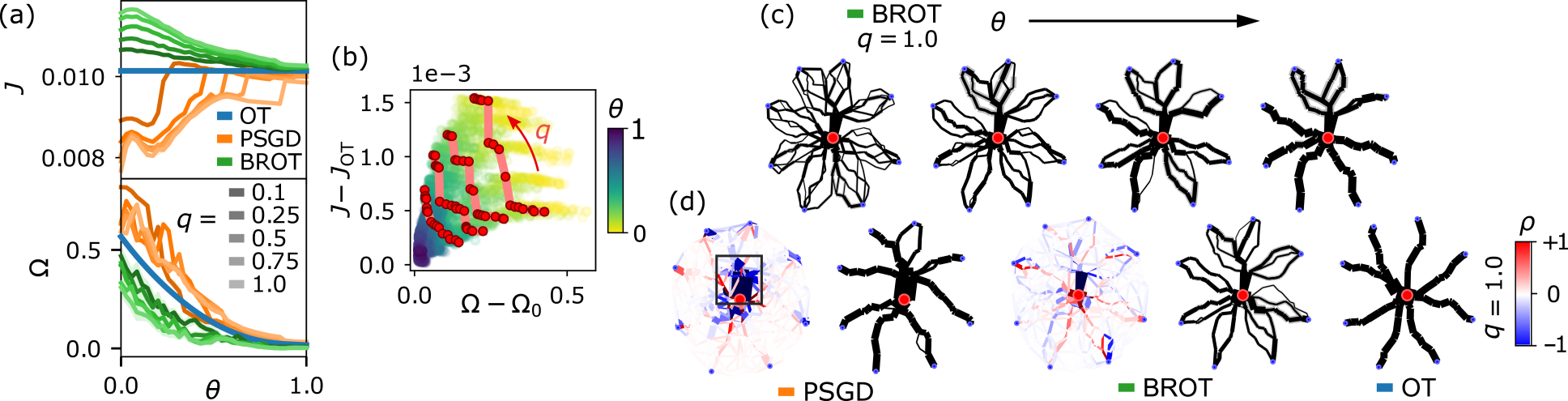}
\caption{Overview of the routing schemes. (a) $J$ and $\Omega$ against $\theta$. (b) Trade-off $J - J_\mathrm{OT}$ vs. $\Omega - \Omega_0$ with varying $(\theta, q,\xi)$. Non-dominated points for $\theta / \theta^{\star} \simeq \{ 0.06, 0.2, 0.3,0.4 \}$ are in red.  (c) BROT's networks at different $\theta$. Edge widths are proportional to the average fluxes in $50$ runs of the algorithm. Gray edge contours are fluxes' standard deviations. (d) Cost (left) and flux (right) networks for all methods and $\theta/\theta^\star=0.4$. Flux networks are as in (c), whereas edges in the cost networks are colored with $\rho$ and their widths are proportional to the fluxes.  The black rectangle frames a region where the network manager triggers high congestion. We conveniently normalize $\theta^{\star}$ and $\rho$.}
\label{fig:panel}
\end{figure*}

\emph{Bilevel Optimization.}---We combine the two optimization problems into one.  Suppose that the network manager is regularly informed of the passengers' routes, and using such information they tune the weights to mitigate traffic.  After each update,  passengers reroute accordingly to the updated weights. 

Formally, this translates into the problem
\begin{align}
\label{eqn:bilevel_opt_1}
&\min_{w \geq \epsilon} \Omega(w; \hat{c}) \\
\label{eqn:bilevel_opt_2}
\mathrm{s.t. } &\; \hat{c} = \argmin_{c \geq 0} J(c;w) \, ,
\end{align}
where the equality in \Cref{eqn:bilevel_opt_2} comes from the convexity of $J$ \cite{bonifaci2012physarum,bonifaci2013short}. In \Cref{eqn:bilevel_opt_1} we explicit the dependence on $w$ as a variable and on $c$ as a parameter (conversely for \Cref{eqn:bilevel_opt_2}).

\emph{Optimal Transport dynamics.}---To find the shortest paths required for the lower-level problem, we couple fluxes and capacities with the ODEs
\begin{alignat}{2}
\label{eqn:dynamics}
\frac{d c_e^i}{d t} &= \frac{{F_e^i}^2}{c_e^i}  - c_e^i \,,
\end{alignat}
where fluxes obey Kirchhoff's law. In \Cref{eqn:dynamics}, edges with high flux enlarge, whereas those where the negative decaying term prevails shrink. Crucially, asymptotic solutions converge to the minimum OT cost $J$ in \Cref{eqn:j_cost_def}, being the Wasserstein distance between passengers' entry and exit distributions \cite{si},  and whose minimizers are origin-destination shortest paths. Beside Kirchhoff's law and positivity, capacities in \Crefrange{eqn:j_cost_def}{eqn:j_cost} are otherwise unconstrained. One can potentially add additional constraints, e.g., a limited budget, by employing recent ideas in the context of adaptation equations \cite{ibrahim2023optimal}. We do not explore this here.

\emph{Projected Stochastic Gradient Descent (PSGD).}---Minimization of \Cref{eqn:bilevel_opt_1} is performed using SGD with a projection step to enforce $w \geq \epsilon$. Importantly, we can derive a closed-form expression for the gradients $\Psi_e = \partial \Omega / \partial w_e$ \cite{si}. To explore the non-convex landscape of the minimization in \Crefrange{eqn:bilevel_opt_1}{eqn:bilevel_opt_2}, we update the weights with dropout at each step, i.e.,  setting to zero $|E|(1-q)$ random gradients, where $0 \leq q \leq 1$. For $q=1$ we get vanilla GD.

\emph{Bilevel optimization scheme.}---In order to find the optimal $c$ and $w$, and hence $F$, we iterate between \Cref{eqn:dynamics} and PSGD recursively. The scheme is repeated until $J$ and $\Omega$ converge. A diagram outlining the optimization method is in \Cref{fig:example_fig}, we also provide an open-source code (\model, Bilevel Routing on networks with Optimal Transport) \cite{git_repo}.

\emph{Experimental setup.}---We analyze BROT's optimal networks against two baselines. The first, referred to as OT, consists of finding passengers' shortest paths without any intervention from the network manager. We assume a unitary cost per unit of length fare, i.e., we set $w = \ell$ with $\ell$ the Euclidean lengths of the edges, and numerically integrate \Cref{eqn:dynamics}.  The second, referred to as PSGD, reflects the scenario of a network manager that tunes $w$ only relying on the shortest paths taken when $w = \ell$, and that disregards how fluxes redistribute while updating $w$. In practice, this corresponds to running PSGD only, with initial conditions being $w(0) = \ell + \xi$ and $c_e^i \simeq |F_{\text{Dij},e}^i|$ \cite{si}, and then,  to integrating \Cref{eqn:dynamics} with $w = w^\star_\text{PSGD}$ being the optimal weights returned by the network manager. Here, $\xi$ is a small zero-sum uniform noise, $F_{\text{Dij}}$ are the shortest path fluxes computed with Dijkstra's algorithm, and the approximation arises because, to avoid numerical instabilities, a small non-zero $c_e^i$ is allocated to all edges. We fix BROT's initial conditions to $w(0) = \ell + \xi$ and $c_e^i (0)= S_{O^i}^i$.

\emph{Synthetic experiments.}---First, we study a network of size $|V| = 300$, $|E| = 864$,  with nodes placed uniformly at random in the unitary disk, and edges extracted from their Delaunay triangulation. Entry and exit inflows are $S^i_{O^i} = +1$ on an origin node at the center, and $S^i_{D^i} = -1 / D$, on $D = 4,8$ destinations $D^i$ on the disk edge. Since $M = 1$, there is only a single index $i$.  Here we discuss results for $D=8$,  for experiments with varying $q$ for $D=4,8$, see Supp. Mat. \cite{si}.

We evaluate $J$ and $\Omega$ at convergence for all methods with different $q$, and ranging $\theta$ from $\theta=0$ to a large value $\theta^\star$ where few edges are congested.  Results are in \Cref{fig:panel}(a). 

Since for OT the network manager does not intervene, $J$ is constant for at all $\theta$, and it is the origin-destination shortest length. Its profile changes when the network manager influences passengers' routes by tuning the weights. Specifically, for PSGD $J$ drops when reducing $\theta$, making it cheaper for the passengers to move. On the contrary, lower $\theta$ corresponds to a larger $J$ for BROT. This behavior seemingly favors an uninformed network manager (PSGD) over an informed one (BROT). However, the profile of $\Omega$ shows that, even though the traveling cost of PSGD is cheaper, all transport networks at convergence are highly congested (large $\Omega$). BROT successfully trades off the cost of traveling against traffic, outputting low values of $\Omega$ for all $\theta$, with only a mild increase as $\theta$ approaches zero.  This is clarified in \Cref{fig:panel}(c), where BROT generates ramified loopy networks. 

The dropout parameter $q$ allows to explore the minimization landscape of  \Crefrange{eqn:bilevel_opt_1}{eqn:bilevel_opt_2}. By decreasing $q$, i.e.,  setting more gradients to zero,  BROT returns lower $J$s at all $\theta$, whereas PSGD gives higher ones,  conversely for $\Omega$. This impacts the network topologies, which are less ramified and akin to OT trees \cite{banavar2000topology}, when $q$ is lower and for the same $\theta$ \cite{si}.  The trade-off between $J$ and $\Omega$ is further laid out in \Cref{fig:panel}(b) where we show $J-J_\text{OT}$ against $\Omega - \Omega_{0}$, $\Omega_{0} = 0$.  We highlight in red the non-dominated points (also referred as maximal points) at four values of $\theta$, computed over all $q$ [as in \Cref{fig:panel}(a)] and $25$ random initializations of BROT. Such points are the best $J$-$\Omega$ trade-off attained by the experimental runs \cite{si}.

For all $q$ and sufficiently low $\theta$, the Price of Anarchy (PoA) \cite{notePoA} is greater for PSGD than for OT, i.e., the network manager's intervention increases traffic congestion, having the opposite effect to that intended. We illustrate exemplary networks at convergence in \Cref{fig:panel}(c). The parameter $\rho = w^\star_X - \ell$ ($X= $ BROT, PSGD), expressing the variation of cost, indicates that the uninformed network manager naively---and significantly---decreases the cost of a small fraction of edges [squared in \Cref{fig:panel}(d)]. This encourages fluxes to largely concentrate on them, thus creating congestion.

To further discern the nature of congestion, we propose two additional metrics.  First, the Gini coefficient of the fluxes, $\text{Gini} = \sum_{mn} | x_m - x_n | / 2 |E|^2 \bar{x}$, where $ \bar{x} = \sum_e  x_e / |E|$ and $x_e = \sum_i |F_e^i|$. $\text{Gini} = 0$ corresponds to uniformly distributed fluxes, and larger $\text{Gini}$ to high congestion. Second, the total travel time
$T_\theta(s) = \sum_{ei} t_{\theta,e}(s) |F_e^i|$, computed with an affine latency function for over-trafficked edges \cite{li2022bilevel, roughgarden2002how}, namely $t_{\theta,e}(s) = \ell_e(1 + s \Delta_e / \theta)/v_\infty$ if $\sum_i |F_e^i| \geq \theta$, and $t_{\theta,e}(s) = \ell_e/v_\infty$ otherwise. Here $v_\infty = 1$ is a (conventionally fixed) free-flow velocity, and $s$ is a sensitivity coefficient to penalize traffic. Results are in \Cref{fig:gini_t}.

The Gini coefficient of PSGD fluctuates slightly around the high values attained by the congested shortest path network of OT.  For BROT, as $\theta$ decreases---more flux gets penalized---Gini sharply drops, yielding progressively distributed networks. The total travel time reveals once again that the uncoordinated action of passengers and network manager may be detrimental compared to having no tuning of $w$. In fact,  times for PSGD are higher than those for OT. BROT keeps $T_\theta(s)$ small for any value of $\theta$ and for both low and high sensitivity.  Finally, as $\theta$ increases, traffic gradually mitigates, with $\lim_{\theta \to + \infty} T_\theta(s) = T_\infty$ ($T_\infty = J_\text{OT}$) being the travel time for infinite capacities, when all passengers flow freely. 

\begin{figure}[t]
\centering
\includegraphics[width=0.91\linewidth]{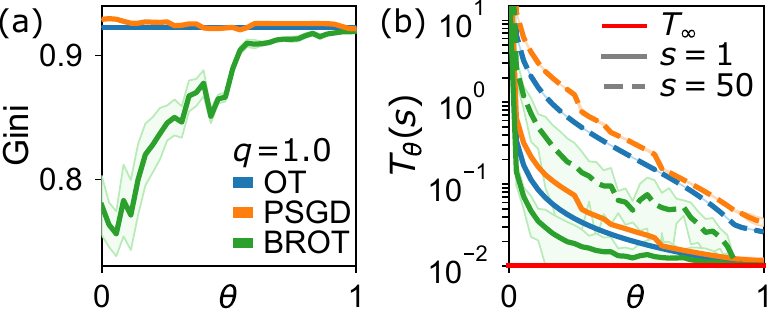}
\caption{Measuring traffic congestion, $D=8$. (a) Gini coefficient against $\theta$.  (b) $T_\theta(s)$ against $\theta$. Solid lines correspond to low sensitivity $s=1$ and dashed ones to $s=50$, in red we draw $T_\infty$ (free flow).  Shades are standard deviations over $50$ realizations of the algorithms.}
\label{fig:gini_t}
\end{figure}

\emph{The E-road network.}---We study the methods on a graph extracted from the International European highways (E-road) \cite{kunegis2013konect,subelj2011robust}, of size $|V| = 541$ and $|E| = 712$. Entry inflows of passengers are populations of $15$ large cities. We assume that all passengers travel from one city to another. Thus, we set for $O^i$ and $v \in D^i$ (being also origin nodes $O^j$) the exiting number of passengers $\tilde{S}_v$ to be proportional to the product $r_v = \tilde{S}_{O^i} \tilde{S}_{O^j}$, properly normalized to ensure conservation of mass. In this way, cities with high inflows have large outflows, and vice versa for small ones. The total number of passengers to be routed is $\sum_i \tilde{S}_{O^i} \simeq 3 \cdot 10^7$.  We fix $\tilde{\theta}$ (dimensionalized by $S_c$) so that $43 \%$ of the passengers reroute from their congested shortest path, found with Dijkstra's and $w = \ell$. 

Results are in \Cref{fig:nets_real}. We observe that in the shortest path configuration of OT,  a large volume of passengers travels between the two most populous cities, Madrid and Berlin, on the southernmost region of the network. The uninformed network (PSGD) heavily increases the price of the connections to Milan \cite{si}. This causes a heavy rerouting from Madrid to the north, and congests the roads connecting Madrid to Paris, and then from Paris to Berlin.  In contrast, BROT distributes traffic over a ramified road network.

\begin{figure}[t]
\centering
\noindent\includegraphics[width=0.925\columnwidth]{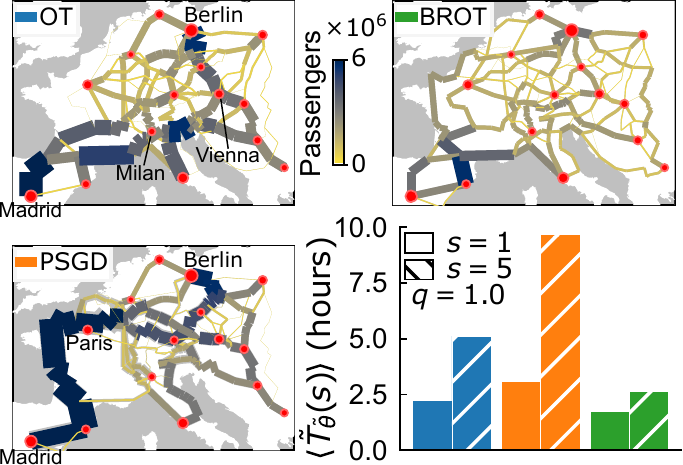} 
\caption{E-road transport networks. Nodes in red are 15 main cities taken as passenger inflows, their size is proportional to the entry inflows. Edge widths are the total number of passengers $\sum_i |\tilde{F}_e^i|$, gray shades are standard deviations over $50$ realizations of the algorithms.}
\label{fig:nets_real}
\end{figure}

We study the average travel time for all routing schemes. This is $\langle \tilde{T}_{\tilde{\theta}}(s) \rangle = \sum_{ei} \tilde{t}_{e,\tilde{\theta}}(s)|\tilde{F}_e^i|/ \sum_{ei} |\tilde{F}_e^i|$, where $\tilde{t}_{\tilde{\theta}}$ is a dimensionalized latency function computed using $\tilde{\ell}$, the Euclidean distance between cities, and $v_\infty = 100 \, (\text{km}/\text{hours})$. 

Results for $s=1,5$ in \Cref{fig:nets_real} show that the average travel time of BROT is substantially lower than that of OT and PSGD. Particularly, for low sensitivity BROT's $\langle \tilde{T}_{\tilde{\theta}}(s) \rangle$ is approximately $1.7$ (hours), while OT's and PSGD's are $2.3$ (hours) and $3.1$ (hours). Here, BROT leads to a reduction in traveled time of approximately $26\%$ and $45\%$ compared to OT and PSGD. This result becomes starker if the sensitivity increases,  here BROT reduces $\langle \tilde{T}_{\tilde{\theta}}(s) \rangle$ of $48\%$ compared to OT---from $5$ (hours) to $2.6$ (hours)---and of $74\%$ compared to PSGD---whose heavy congestion gives $\langle \tilde{T}_{\tilde{\theta}}(s) \rangle \simeq 10$ (hours).  Once again, the PoA (the travel time) is higher if the network manager's intervention is uncoordinated with the passengers (PSGD), as opposed to when there is no intervention (OT).

Experiments on the E-road network for $q=0.25,0.5,0.75$ are in Supp. Mat. \citep{si}.

\emph{Conclusion.---}BROT relies on theoretical assumptions that can be challenging to meet in real-world traffic control \cite{papageorgiou2003review}, e.g., passengers rerouting more unpredictably than expected by theoretical models. Nevertheless, our analysis on the E-road network demonstrates how an informed tuning of road tolls---where the network manager factors in passengers' rerouting---can be beneficial for reducing the carbon footprint of roads, since traffic jams, and hence longer travels, critically impact greenhouse gas emissions of vehicles \cite{van2001managing, kellner2016exploring, barth2009traffic}. 

To facilitate practitioners using our algorithms, we open-source our code \cite{git_repo}.

\emph{Acknowledgments.}---The authors thank the International Max Planck Research School for Intelligent Systems (IMPRS-IS) for supporting Alessandro Lonardi. 

\bibliography{bibliography}

\begin{thebibliography}{56}
\bibitem [{not()}]{notePoA}%
  \BibitemOpen
  \href@noop {} {}\bibinfo {note} {The price of anarchy measures how much a
  system degrades due to greedy behavior of its agents. In our case, the global
  efficiency of the system is measured by $\Omega$, which increases as the
  network gets more congested by greedy passengers.}\BibitemShut {Stop}%
\end{thebibliography}


\onecolumngrid
\mbox{}
\clearpage
\newpage

\setcounter{equation}{0}
\setcounter{figure}{0}
\setcounter{section}{0}
\setcounter{table}{0}
\setcounter{page}{1}
\makeatletter
\renewcommand{\theequation}{S\arabic{equation}}
\renewcommand{\thefigure}{S\arabic{figure}}
\renewcommand{\thetable}{S\arabic{table}}

\title{Bilevel Optimization for Traffic Mitigation in Optimal Transport Networks: Supplementary Material (SM)}

\maketitle
\onecolumngrid\vspace{-1.3cm}

\section{Problem formulation}

When building the model setup, we assume that passengers move greedily according to Wardrop's first principle. This means that, for a given set of weights $w$,  they travel from their origins $O^i$ to their destinations $D^i$ minimizing the OT cost $J = \sum_{ei} w_e |F_e^i|$. As a consequence, a meaningful comparison of BROT is proposed against two methods. The first, PSGD, consists of a scheme where initially only $\Omega$ is minimized by a network manager that tunes $w$ while ignoring how passengers reroute when the weights update. We suppose that, at $t=0$, the network manager knows which would be the paths that the passengers were to take if they moved minimizing $J$ with $w = \ell$---we compute these fluxes with Dijstra's algorithm. In PSGD, only at convergence of $w$ the passengers choose on which path to travel. The second scheme is OT, where the passengers find their shortest path for $w = \ell$. Here, the network manager does not intervene on $w$.

\begin{figure*}[htpb]
\centering
\includegraphics[width=0.63\linewidth]{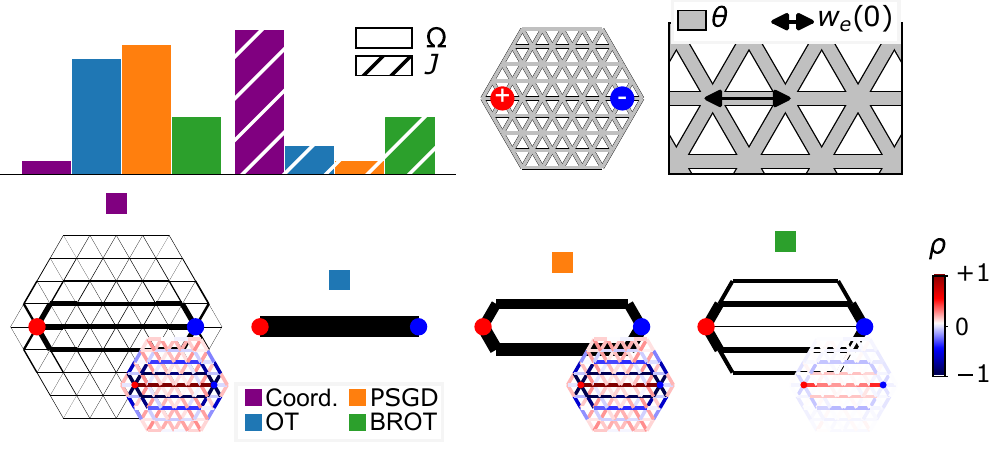}
\caption{Bilevel optimization scheme in detail. We plot costs and network topology for different adaptation rules. Histograms and networks labeled with purple, blue, orange, and green squares correspond to coordinated traffic, OT, PSGD, and BROT, respectively. Hatch styles are $J$ and $\Omega$.  Edges in black are proportional to $\sum_i |F_e^i|$, while those colored with $\rho$ express the change in (the normalized) $w$.}
\label{fig:explanatory_supp}
\end{figure*}

We give an overview of the schemes in \Cref{fig:explanatory_supp}, which is a companion Figure of Fig. 1 (main text). Here we consider an entry and an exit node where unitary mass flows in (red plus) and flows out (blue minus) the network. Additionally to BROT, PSGD (with dropout probability fixed at $q=1$, i.e., vanilla GD), and OT, we also extract the paths the passengers would take if they did not act greedily. This simulates the unrealistic---for our case of study---situation where the agents on a road network follow the instruction of an oracle, and accept to reroute in order to achieve a social optimum rather than maximizing their own benefit.

The costs $J$ and $\Omega$ in \Cref{fig:explanatory_supp} show that if passengers unrealistically moved in a coordinated way, then traffic congestion would be greatly minimized. However, this would cause $J$ to explode.  The OT cost $J$ can be minimized by OT and PSGD, which in turn trigger traffic congestion. As discussed in the main text,  it may happen that congestion is larger in PSGD than in OT, showing that the uncoordinated actions of passengers and network manager can be detrimental to the global efficiency of the transport network. Only BROT is able to trade off between $J$ and $\Omega$.

The networks in \Cref{fig:explanatory_supp} reflect the costs profiles. Particularly, in the coordinated transport network (purple), passengers distribute over the whole area of the network and keep congestion low. Here, $\rho = w^\star_{\text{PSGD}} - \ell$ ($w^\star_{\text{PSGD}}$ are the weights tuned by the network manager) shows that the straight line connecting the origin and destination nodes is heavily penalized,  while other parallel connections and oblique links branching from the origin and destination become cheaper. In the OT network (blue), passengers simply travel on the shortest straight path.  In PSGD (orange), passengers split in two separate branches, which however are congested. Only BROT (green) is able to trade off between congestion and total travel cost by outputting a network where fluxes distribute hierarchically to prevent over-trafficking and to minimize the travel cost.  Noticeably, the action of the network manager (here quantified by $\rho = w^\star_{\text{BROT}} - \ell$) is less invasive in this scenario, i.e., edge costs need to be varied less to find an optimal configuration.

\section{Nondimensionalization of the model}

The scale-independent adaptation equations for the evolution of weights and capacities can be derived by rescaling dimension-dependent quantities.  We start with the constant coefficients ODEs
\begin{align}
\label{eqn:dimensional_ode_ot}
\frac{d\tilde{c}^i_e }{d\tilde{t}} &= \alpha \frac{\tilde{F}^{i2}_e }{\tilde{c}^i_e} - \beta \tilde{c}^i_e \\
\label{eqn:dimensional_ode_gd}
\frac{d\tilde{w}_e}{d\tilde{t}} &= - \gamma \frac{ \partial \tilde{\Omega}(\tilde{w}(t),\tilde{c})}{ \partial \tilde{w}_e} \,.
\end{align}
In \Cref{eqn:dimensional_ode_ot} we write a dimension-dependent version of Eq. (8, main text). In \Cref{eqn:dimensional_ode_gd} the gradient flow equation associated with the GD update used to tune the weights. Here, $\alpha$, $\beta$, and $\gamma$ are constant coefficients with appropriate dimensions. We then choose the following nondimensionalization:
\begin{align}
\label{eqn:dim}
t &= \tilde{t} / t_c \\
c_e &= \tilde{c}_e / c_c \\
w_e &= \tilde{w}_e / w_c \\
\label{eqn:dim1}
S_v &= \tilde{S}_v / S_c
\end{align}
where $t_c, c_c, w_c$ and $S_c$ are characteristic units.  As mentioned in the main text, in order to fix ideas, one could think of such units as time, length, price of tolls applied to roads, and number of passengers, for $t_c$, $c_c$, $w_c$, and $S_c$, respectively.  Importantly, such a choice of the parameters' dimensions is not binding since for different physical scenario the problem variables assume diverse meanings. For example, in the case of plant leaves the capacity $c$ could be thought of as the production rate of auxin transporting proteins \cite{rolland2005reviewing,ronellenfitsch2019phenotypes}. Substituting \Crefrange{eqn:dim}{eqn:dim1} in Kirchhoff's law yields $F_e = \tilde{F}_e/S_c$, where $F$ are nondimensional fluxes.  By recasting all nondimensional variables in \Cref{eqn:dimensional_ode_ot} and  \Cref{eqn:dimensional_ode_gd} we get
\begin{alignat}{2}
    \frac{d {c_e^i} }{d {t}} &= \alpha \left( \frac{t_c S_c^2}{c_c^{2}} \right)\frac{F_e^2}{c_e^i} - \beta t_c \, {c_e^i} \\
\frac{d {w}_e}{d {t}} &= - \gamma \left( \frac{t_c S_c^2}{w_c^2} \right) \frac{ \partial {\Omega} ({w(t)}, {c})}{ \partial {w}_e}
\end{alignat}
showing that, to recover the adimensional model, we can conventionally set
\begin{align}
t_c &= 1/b \\
c_c/S_c &= \sqrt{a/b}\\
{w_c}/{S_c} &= \sqrt{c/b}\,.
\end{align}
We also notice that $\tilde{J} = \sum_{ei} \tilde{w}_e | \tilde{F}_e^i | = (w_c S_c) \sum_{ei} {w}_e | {F}_e^i  | $, which is $\tilde{J} = (w_cS_c) J$. Moreover, $\Omega = \tilde{\Omega} / S_c^2$, with $\theta = \tilde{\theta}/S_c$ for an opportunely dimension-dependent capacity threshold $\tilde{\theta}$, and $\tilde{T}(s) = (w_{\text{OT},c} S_c) T(s)$, where $w_{\text{OT},c}$ is the nondimensionalization coefficient used for the constant weights of OT. As a consequence, when comparing the costs $\Omega$ and $J$, and the total traveled time $T_{\theta}(s)$ between different methods we perform the following nondimensionalization.   We fix $w_{{X},c} = \sum_e \tilde{w}^\star_{{X},e} $ where ${X}$ is one between OT, BROT, or PSGD, and starred weights are those at convergence. Additionally, we set $S_c$ to be the total inflowing number of passengers, i.e., $S_c =  \sum_{i} \tilde{S}_{O^i}^i$. This yields, for any algorithm $X$,
\begin{align}
J_{X} = \frac{\tilde{J}_{X}}{\left(\sum_e \tilde{w}^\star_{{X},e} \right) \left( \sum_{i} \tilde{S}_{O^i}^i \right) } \quad \Omega_{X} = \frac{\tilde{\Omega}_{X}}{ \left( \sum_{i} \tilde{S}_{O^i}^i \right)^2 }  \quad T_{\theta,X}(s) = \frac{\tilde{T}_{\tilde{\theta},X}(s)}{\left(\sum_e \tilde{w}^\star_{{\text{OT}},e} \right) \left( \sum_{i} \tilde{S}_{O^i}^i \right) } \,.
\end{align}

\section{Connection with Optimal Transport}

The adaptation rules governing the evolution of the capacities [Eq. (8, main text)] is tightly connected with Optimal Transport (OT) theory. In detail, consider the problem formulation proposed in the main text. By fixing $i = 1$ we obtain that the passengers inflows $\mu = S_{O^i}$ and outflows $\nu = - S_{D^i}$ are two atomic probability distributions supported on the origin and destination nodes, respectively, that need to be mapped one into the other by minimizing the transportation cost. This problem, using a standard OT formulation (primal Kantorovich Problem) is the linear program \cite{kantorovich1960mathematical}
\begin{align}
\label{eqn:kantorovich_problem}
\min_{\pi \in \Pi(\mu,\nu)} \sum_{u \in V, v \in V} {w_{uv} \pi_{uv}} \,.
\end{align}
Here $\Pi(\mu,\nu)$ is the set of transport paths $\pi$ (expressing the probability of an assignment of passengers on $uv$) that satisfy the conservation constraints $\sum_v \pi_{uv} = \mu_u$ and $\sum_u \pi_{uv} = \nu_v$. The cost $w_{uv}$ corresponds to the price one needs to pay to move passengers from $u$ to $v$. Since the transport of passengers is supported on a network, we fix $w_{uv} = +\infty$ if two nodes $u,v$ are not connected. It can be proved \cite{lonardi2023immiscible} that the problem in \Cref{eqn:kantorovich_problem} and the minimization problem in Eq. (2, main text) correspond, i.e., the minimization objectives and the search space defined by $\Pi(\mu,\nu)$ and Kirchhoff's law are the same, therefore shortest path fluxes are exactly OT paths. Moreover, OT fluxes are asymptotic solutions of Eq. (8, main text) \cite{lonardi2023immiscible}.

Noticeably, the optimal cost $J$ is exactly the Wasserstein distance between $\mu$ and $\nu$, when $w_{uv}$ satisfies the properties of a metric: (i) symmetry, (ii) vanishing along the diagonal, (iii) triangular inequality \cite{villani2008optimal}.  These requirements may not hold in our setup, nevertheless, $J$ can still be interpreted as the cost that passengers pay to move on the network. We also remark that even if disconnected nodes yield an infinite edge cost, the finiteness of \Cref{eqn:kantorovich_problem} is guaranteed by assuming the existence of at least one path joining $O^i$ and $D^i$ where passengers can travel.

OT paths for $ i > 1 $ are the overlap of $M$ independent solutions of \Cref{eqn:kantorovich_problem},  with passengers that move from $\mu^i = S_{O^i}$ to $\nu^i = - S_{D^i}$ for all $i$.

\section{Closed-form expression of the upper-level problem gradients}

We derive the gradients of the upper-level problem in Eq. (6, main text) in closed-form.  Let us start by applying the chain rule to $\Omega$, which is defined as
\begin{align}
\Omega = \frac{1}{2} \sum_e \Delta_e^2 H(\Delta_e) \,,
\end{align}
with $\Delta_e = \sum_i |F_e^i| - \theta$ and $H$ Heaviside step function.  In order to ease notations, for any feasible set of capacities $c$, we denote all edges for which $\Delta_e \geq 0$ as $w_e \in W$.  Conversely,  if $\Delta_e < 0$ (hence $H(\Delta_e) = 0$), then $w_e \notin W$.  Moreover, we compactly use $||F_e||_1 = \sum_i |F_e^i|$. This allows us to write
\begin{align}
\Psi_e &= \frac{\partial \Omega}{\partial w_e} \\ &= \sum_{e' \in W} \frac{\partial \Omega}{\partial \Delta_{e'}} \frac{\partial \Delta_{e'}}{\partial w_e} \\
\label{eqn:chain_rule_1}
&=\sum_{e' \in W} \Delta_{e'} \frac{\partial ||F_{e'}||_1}{\partial w_e} \,.
\end{align}
To manipulate \Cref{eqn:chain_rule_1}, we introduce the auxiliary variables ${\Res}^i_\Iedge = w_e / c_e^i$,  that we can use to write compactly all problem's main variables. In particular, we denote fluxes, as variables of $\Res$, as $\mathcal{F}^i_e =  (\mathcal{P}^i_u - \mathcal{P}^i_v) / {r^i_e}$. Least-squares potentials are $\mathcal{P}^i_v = \sum_u (\mathcal{L}^{i\dagger})_{vu} S_u^i$, and entries of the Laplacian are $\mathcal{L}^i_{vu} =\sum_{e} (1/ \Res^i_e) \Inc_{ve} \Inc_{ue}$. With this change of variable, derivatives of the fluxes in \Cref{eqn:chain_rule_1} become
\begin{align}
\label{eqn:partial_der_substituteinto}
\frac{\partial ||F_{e'}||_1}{\partial w_e} = \sum_i \frac{\partial ||\mathcal{F}_{e'}||_1}{\partial r^i_e} \frac{\partial r^i_e}{\partial w_e} \, .
\end{align}
Now, write differences of pressure along the network edges as $\Delta \mathcal{P}^i_e = \mathcal{P}^i_u - \mathcal{P}^i_v$, and when computing the gradients with respect to an edge $e$, we separate contributions for $e' \neq e$, and for $e' = e$. In particular, we write
\begin{align}
\frac{\partial ||\mathcal{F}_{e'}||_1}{\partial r^i_e} &= \frac{\partial}{\partial r_e^i} \left( \sum_j \frac{1}{r_{e'}^j} | \Delta \mathcal{P}_{e'}^j |  \right)\\
\label{eqn:partial_der_apx_last_equal}
&= 
\displaystyle\begin{cases}
\displaystyle\frac{1}{r_{e'}^i} \text{sgn}({\mathcal{P}_{e'}^i}) \frac{\partial \Delta \mathcal{P}_{e'}^i}{\partial r_e^i} \quad &\forall e' \neq e \\
- \displaystyle\frac{1}{r_{e'}^{i2}} | \Delta  \mathcal{P}_{e'}^i |  + \frac{1}{r_{e'}^i} \text{sgn}({\mathcal{P}_{e'}^i}) \frac{\partial \Delta \mathcal{P}_{e'}^i}{\partial r_{e'}^i} \quad &\forall e' = e \, .
\end{cases}
\end{align}
To simplify notations, we substitute \Cref{eqn:partial_der_apx_last_equal} into \Cref{eqn:partial_der_substituteinto}, and group all terms in one unique expression. This reads
\begin{align}
\label{eqn:compact_der}
\frac{\partial ||F_{e'}||_1}{\partial w_e} = \sum_i \bigg( \frac{1}{r_{e'}^i} \text{sgn} (\Delta  \mathcal{P}_{e'}^i ) \frac{\partial \Delta \mathcal{P}_{e'}^i}{\partial r_e^i}  - \frac{1}{r_{e'}^{i2}} | \Delta  \mathcal{P}_{e'}^i | \delta_{e' e} \bigg) \frac{\partial r_e^i}{\partial w_e}
\end{align}
with $\delta_{e'e}$ Kronecker delta.
Moreover, derivatives of pressure differences can be written making explicit the definition of the least-squares potential, that is,
\begin{align}
\label{eqn:partial_der_apx}
\frac{\partial \Delta \mathcal{P}_{e'}^i}{\partial r_e^i} =\sum_{\Inode \Inodetwo} \Inc_{\Inode \Iedgetwo} \f{\partial (\mathcal{\Lap}^{i\dagger})_{vu}}{\partial {\Res^i_e}} \NetMass_\Inodetwo \,.
\end{align}

Now, in order to conclude the derivations, we need to calculate the derivatives of the Laplacian Moore-Penrose inverse in \Cref{eqn:partial_der_apx}. This can be done by following the detailed calculation of \Cref{secAPX:moorepenrose}.  We denote their closed-form expression with $ \mathcal{Q}^i_{\Inodetwo \Inode \Iedge}  = {\partial (\mathcal{\Lap}^{i\dagger})_{\Inode \Inodetwo}} / {\partial {\Res}^i_\Iedge}$ as defined in \Cref{eqnAPX:derivative_3}.

Substituting \Cref{eqnAPX:derivative_3} into \Cref{eqn:partial_der_apx}, and then \Cref{eqn:partial_der_apx} back into \Cref{eqn:compact_der}, we obtain
\begin{align}
\label{eqnAPX:chain_rule}
\frac{\partial || F_{e'}||_1}{\partial w_e} =  \sum_i \left( \frac{1}{r_{e'}^i} \text{sgn}(\Delta  \mathcal{P}_{e'}^i ) \sum_{vu} B_{ve'} \mathcal{Q}^i_{vue}S_u^i  -  \frac{1}{r_{e'}^{i2}} | \Delta  \mathcal{P}_{e'}^i | \delta_{e'e} \right) \frac{\partial r_e^i}{\partial w_e} \,.
\end{align}
Making explicit capacities and weights in all terms of \Cref{eqnAPX:chain_rule} yields
\begin{alignat}{2}
\frac{\partial || {F}_{e'}||_1}{\partial w_e} &=  \sum_i \bigg( \frac{c_{e'}^i c_e^i}{w_{e'} w_e^2}  \text{sgn}(\Delta  P_{e}^i )\sum_{vu} B_{ve'} \Lambda^i_{vue} S_u^i  -  \frac{c_{e'}^i}{w_{e'}^2}   | \Delta  {P}_e^i | \delta_{e'e} \bigg) \quad &&\forall e \in E, \, e' \in W \\
\label{eqn:partial_der_toplug}
\Lambda^i_{vue} &=  {\Lap^{i\dagger}_{u x}} {\Lap^{i\dagger}_{x v}} + {\Lap^{i\dagger}_{u y}} {\Lap^{i\dagger}_{y v}} -  {\Lap^{i\dagger}_{u y}} {\Lap^{i\dagger}_{x v}} -  {\Lap^{i\dagger}_{u x}} {\Lap^{i\dagger}_{y v}} \quad &&\forall e = (x,y) \in W,  \, u,v \in V,  \, i \in M\\
\Delta P^i_e &= p^i_{x} - p^i_{y}&& \forall e = (x,y) \in W,  \, i \in M \,,
\end{alignat}
which we simplify further by writing
\begin{align}
\frac{\partial || F_{e'} ||_1}{\partial w_e} &= \sum_i \bigg( \frac{c_{e'}^i c_e^i}{w_{e'} w_e^2}  \text{sgn}(\Delta  P_{e'}^i )  \sum_{vu} B_{ve'} \Lambda^i_{vue} S_u^i - \frac{|F_{e'}^i|}{w_{e'}} \delta_{e'e}  \bigg) \\
&= \sum_i \bigg( \frac{c_{e'}^i c_e^i}{w_{e'} w_e^2}  \text{sgn}(\Delta  P_{e'}^i ) \sum_{vu} B_{ve'} ({L^{i\dagger}_{u x}} {L^{i\dagger}_{x v}} + {L^{i\dagger}_{u y}} {L^{i\dagger}_{y v}} -  {L^{i\dagger}_{u y}} {L^{i\dagger}_{x v}} -  {L^{i\dagger}_{u x}} {L^{i\dagger}_{y v}}) \NetMass_u - \frac{|F_{e'}^i|}{w_{e'}} \delta_{e'e}  \bigg) \\
&= \sum_i \bigg( \frac{c_{e'}^i c_e^i}{w_{e'} w_e^2}  \text{sgn}(\Delta  P_{e'}^i )  \sum_{vu} B_{ve'} ({L^{i\dagger}_{v{x}}} p_{x} + {L^{i\dagger}_{v{y}}}p_{y} - {L^{i\dagger}_{v{x}}} p_{y} - {L^{i\dagger}_{v{y}}} p_{x}) - \frac{|F_{e'}^i|}{w_{e'}} \delta_{e'e}  \bigg) \\
&= \sum_i \bigg( \frac{c_{e'}^i c_e^i}{w_{e'} w_e^2} \text{sgn}(\Delta  P_{e'}^i )   \sum_{vu} B_{ve'} \Inc_{u \Iedge} {L^{i\dagger}_{vu}} \Delta P_e^i  - \frac{|F_{e'}^i|}{w_{e'}} \delta_{e'e}  \bigg) \\
&= \sum_i \bigg( \frac{c_{e'}^i}{w_{e'}} \frac{F^i_e}{w_e}  \text{sgn}(F_{e'}^i )  \sum_{vu} B_{ve'} \Inc_{u \Iedge} {L^{i\dagger}_{vu}} - \frac{|F_{e'}^i|}{w_{e'}} \delta_{e'e}  \bigg) \\
\label{eqn:last_expr_chain}
&= \sum_i  \frac{F_e^i}{w_e}  \bigg( \frac{c_{e'}^i}{w_{e'}} \text{sgn}(F_{e'}^i ) G_{e'e}^i - \mathrm{sgn}(F_e^i)  \delta_{e'e}  \bigg)  \,,
\end{align}
where we introduced $G^i_{e'e} = \sum_{vu} B_{ve'}B_{ue} {L^{i\dagger}_{vu}} $. Notice that $G$ yields non-zero terms for all edges in $W$ and not, hence, congestion on an edge $e$ can affect the weight of any other edge of the network. We conclude by substituting \Cref{eqn:last_expr_chain} in \Cref{eqn:chain_rule_1}, and finally obtain the gradients of $\Omega$ in closed-form. These read
\begin{align}
\label{eqn:final_gradient}
\Psi_e = \frac{\partial \Omega}{\partial w_e} = \sum_{e' \in W} \Delta_{e'} \sum_{i}  \frac{F_e^i}{w_e}  \bigg( \frac{c_{e'}^i}{w_{e'}} \mathrm{sgn} ( F_{e'}^i ) G_{e'e}^i - \mathrm{sgn}(F_e^i) \delta_{e'e}  \bigg) \,.
\end{align}
A similar result can be found in \cite{li2022bilevel} (Appendix IIIC).

\subsection{Differentiation of the Laplacian Moore-Penrose inverse}
\label{secAPX:moorepenrose}

The following calculations can be carried out identically for any index $i \in M$, thus we omit it. We also assume that $\mathcal{\Lap}$ has constant rank, i.e., the network does not disconnect in multiple connected components. With this assumption, we can write (\cite{golub1973differentiation}, Theorem 4.3) the Laplacian Moore-Penrose inverse derivatives as
\begin{align}
\label{eqn:penrose_first_der}
	\f{\partial \mathcal{\Lap}^\dagger}{\partial {\Res}_\Iedge} &= -\mathcal{\Lap}^{\dagger} \frac{\partial \mathcal{\Lap}}{\partial {\Res}_\Iedge} \mathcal{\Lap}^{\dagger} +\mathcal{\Lap}^{\dagger} \mathcal{\Lap}^{\dagger \top} \frac{\partial \mathcal{\Lap}^{\top}}{\partial {\Res}_\Iedge} \left(I-\mathcal{\Lap} \mathcal{\Lap}^{\dagger}\right)+\left(I-\mathcal{\Lap}^{\dagger} \mathcal{\Lap}\right) \frac{\partial \mathcal{\Lap}^{\top}}{\partial {\Res}_\Iedge } \mathcal{\Lap}^{\dagger\top} \mathcal{\Lap}^{\dagger} \,,
\end{align}
where we introduced $I$, identity matrix of size $\Nnode$, and $\top$ denotes the transposed of a matrix. The expression in \Cref{eqn:penrose_first_der} can be combined with the network Laplacian properties $\mathcal{\Lap} \mathcal{\Lap}^\dagger = \mathcal{\Lap}^\dagger \mathcal{\Lap} = I - \vec{1} \otimes \vec{1} / {\Nnode}$, and $\sum_{\Inodetwo} {\partial \mathcal{\Lap}_{\Inodetwo \Inode}} / {\partial{{\Res}_\Iedge}} = \sum_{\Inode}  { \partial \mathcal{\Lap}_{\Inodetwo \Inode}} / {\partial{{\Res}_\Iedge}} = 0$, with $\vec{1}$ being a $\Nnode$-dimensional array of ones and $\otimes$ the Kronecker product. This yields
\begin{align} \label{eqnAPX:derivative_2}
	\f{\partial \mathcal{\Lap}^\dagger}{\partial {\Res}_\Iedge} = -\mathcal{\Lap}^{\dagger} \frac{\partial \mathcal{\Lap}}{\partial {\Res}_\Iedge} \mathcal{\Lap}^{\dagger} \,,
\end{align}
which we further simplify, for all $\Inodetwo, \Inode \in \Nodes$ and $ \Iedge = (x,y) \in W$, as follows:
\begin{align}
	\f{\partial \mathcal{\Lap}^\dagger_{\Inodetwo \Inode}}{\partial {\Res}_\Iedge} &=  - \sum_{u' v' \Iedgetwo} \mathcal{\Lap}^\dagger_{u' u} \f{\partial}{\partial {\Res}_\Iedge} \left( \f{\Inc_{u' \Iedgetwo} \Inc_{v' \Iedgetwo}}{{\Res}_\Iedgetwo} \right) \mathcal{\Lap}^\dagger_{v' \Inode} \\
	\label{eqnAPX:derivative_3}
	&= \f{1}{\Res_\Iedge^2} \left( \mathcal{\Lap}^\dagger_{\Inodetwo x} \mathcal{\Lap}^\dagger_{x \Inode} + \mathcal{\Lap}^\dagger_{\Inodetwo y} \mathcal{\Lap}^\dagger_{y \Inode} -  \mathcal{\Lap}^\dagger_{\Inodetwo y} \mathcal{\Lap}^\dagger_{x \Inode} -  \mathcal{\Lap}^\dagger_{\Inodetwo x} \mathcal{\Lap}^\dagger_{y \Inode} \right) \\ &= \mathcal{Q}_{\Inodetwo \Inode \Iedge} \,.
\end{align}

\section{Numerical implementation}
\subsection{Projected Stochastic Gradient Descent}
\label{ssec:PSGD}

The bilevel minimization problem in Eqs. (6)-(7) (main text) is highly non-convex. This implies that the energy landscape of $(\Omega,J)$ has several local minima, that we wish to explore. Such a task is carried out by implementing batch GD. Particularly, at every discrete time step $n \in \mathbb{N}_0$ we draw an $|E|$-dimensional vectors with components sampled from a Bernoulli probability distribution with parameter $q$, i.e., $\alpha_e(n) \sim \mathrm{Bernoulli}(q)$. We multiply $\alpha(n)$ and $\Psi(n)$ so that, on average, at every step $|E|(1-q)$ random gradients are set to zero.

Additionally, in order constrain the weights onto their feasibility set $C  = \{ w \in \mathbb{R}^{|E|} \, : \,w_e \geq \epsilon > 0 \}$, we perform a projection step at every descent iteration.  Thus, for every discrete time step $n \in \mathbb{N}_0$, we modify the update of the weights as
\begin{alignat}{2}
\text{GD:} \; &w_e(n+1) = w_e(n) - \eta \Psi_e(n) \\
\label{eqn:PSGD}
\text{PSGD:} \; &w_e(n+1) = \argmin_{x_e \geq \epsilon} | (w_e(n) - \eta \alpha(n) \Psi_e(n)) - x_e | \,.
\end{alignat}
Other projection methods could be used to constrain the weights. Particularly, in our numerical code \cite{git_repo} we also implement the method of \citet{muehlebach2022constraints}. This consists of adding a ``cleverly-designed'' momentum term to the descent scheme, which ensures that, given that $w(0) \in C$, then $w$ will fall in the feasibility region at convergence. Remarkably, this approach has the advantage of being easily adaptable to highly non-linear constraints and has recently been used to formulate a constrained dynamical formulation of OT like the one used in this work \cite{ibrahim2023optimal}. However, since the structure of $C$ in our case is simple, we observe that it does not give any numerical benefit compared to PSGD. Hence, we opt for the latter. The threshold $\epsilon$ has been set to $\epsilon = 0.01 \cdot \min w$.

\subsection{Initial conditions}

Initial conditions for OT, PSGD, and BROT are set as follows:
\begin{alignat}{2}
\label{eqn:label_ot_init}
\text{OT:} \;  &c_e^i(0) = S_{O^i}^i &&w_e = \ell_e \\
\label{eqn:label_PSGD_init}
\text{PSGD:} \; &c_e^i = (1 - \lambda)|F^i_{\text{Dij},e}| + \lambda S_{O^i}^i \quad &&w_e(0) = \ell_e + \xi_e \\
\label{eqn:label_brot_init}
\text{BROT:} \; &c_e^i(0) = S_{O^i}^i   &&w_e(0) = \ell_e + \xi_e \,,
\end{alignat}
where $\lambda = 0.95$,  $F_{\text{Dij}}$ are the shortest path fluxes computed with Dijkstra's algorithm, and $\xi$ is zero-sum small noise, defined as $\xi = 0.1 \cdot  \min w(0)( \zeta_0 / \max | \zeta_0 |)$ where $\zeta_0$ the correspondent mean-centered vector of $\zeta$, $\zeta_e \sim U(0,1)$. 

Conditions in \Cref{eqn:label_ot_init} allow us to compute the shortest path fluxes without the intervention of the network manager, hence with $w=\ell$. In this case, the width of roads at $t=0$ is set to be uniform and equal to the inflowing passengers $S_{O^i}^i$, since passengers could potentially travel on any edge of the network.  In PSGD, we suppose that the network manager is---\emph{only initially}---informed about passengers' shortest paths, hence we should initialize the capacities as $c_e^i = |F_e^i|$. This condition comes from the fact that at convergence of Eq. (8, main text) and for optimal solutions of Eq. (2, main text) the scaling $c_e^i \sim |F_e^i|$ holds \cite{lonardi2021designing}. In order to prevent numerical instabilities, in \Cref{eqn:label_PSGD_init} we assign $c_e^i \simeq |F^i_{\text{Dij},e}|$ to edges that are traversed by shortest path fluxes, and $c_e^i = 0.05 \cdot S_{O^i}^i$ to all the others. The weights are initialized as equal to the lengths, with a small zero-sum noise used to explore the cost landscape of $\Omega$, together with the dropout coefficient $\alpha$. In PSGD, $c$ gets updated only after the full update of the weights by the network manager is performed, then the OT paths are computed for $w = w^\star_{\text{PSGD}}$. For BROT, the network manager is initially uninformed about passengers' routes, hence $c_e^i(0) = S_{O^i}^i$. Similarly to PSGD, $w(0) = \ell + \xi$.

\subsection{Implementation details}
\label{ssec:additional_implementation_details}

In \Cref{algo:methods} we write a pseudocode for the implementation of OT, PSGD, and BROT. In \Cref{tab:parameters}, we provide a detailed list of all parameters used for our experiments. Ranging the seeds for $\xi$ and using the random dropouts $\alpha$, we explore the cost landscape of the bilevel optimization problem.

\begin{algorithm}[htpb]
\DontPrintSemicolon
\caption{BROT, PSGD, and OT}
{\textbf{Input:} network $G$, inflows $S$, $\theta$, additional parameters as in \Cref{tab:parameters}.}\\
{\textbf{Initialize:} {\color{otcolor}{OT}}: $w$, $c(0)$ with \Cref{eqn:label_ot_init}; {\color{PSGDcolor}{PSGD}}: $w(0)$, $c$ with \Cref{eqn:label_PSGD_init}; {\color{brotcolor}{BROT}}: $w(0)$, $c(0)$ with \Cref{eqn:label_brot_init}}\\
\While{{\normalfont{convergence is False}}}{
{\color{otcolor}{OT}}: update $c$ with Eq. (8, main text), $w^\star_{\text{OT}} = \ell$ \tcp*{\normalfont{The weights remain fixed until convergence}}
{\color{PSGDcolor}{PSGD}}: update $w$ with \Cref{eqn:PSGD}, at convergence fix $w = w^\star_{\text{PSGD}}$ \tcp*{\normalfont{The capacities remain fixed until convergence}}
{\color{white}{PSGD:}} update $c$ with Eq. (8, main text) \tcp*{\normalfont{The weights are now fixed}}
{\color{brotcolor}{BROT}}: alternate Eq. (8, main text) and \Cref{eqn:PSGD}
}
{\textbf{Output:} optimal weights, capacities, and fluxes $w^\star$, $c^\star$, $F^\star(c^\star, w^\star)$}
\label{algo:methods}
\end{algorithm}

\begin{table}[htpb]
    \centering
    \begin{tabular}{llcccccc} \toprule
    & Network configuration & \multicolumn{5}{c}{Parameters} \\
    & & $\theta$ & $q$ & $T$ & $\varepsilon_{J}$ & $\varepsilon_{\Omega}$ \\
    & & & (dropout prob.) & (num. of iterations) & (conv. thres. of $J$) & (conv. thres. of $\Omega$) \\
    \midrule
    &Lattice exps & range(0.0, 0.505, 0.005) & 1 & $5 \cdot 10^3$ & $10^{-6}$ & $10^{-6}$ \\
    \midrule
    &Synth. exps main text  \\
    &  $D = 4$ & range(0.0, 0.13, 0.005) & $\{ 0.1, 0.25, 0.5, 0.75, 1 \} $ & $5 \cdot 10^3$ & $10^{-6}$ & $10^{-6}$ \\
    &  $D = 8$ & range(0.0, 0.072, 0.002) & $\{ 0.1, 0.25, 0.5, 0.75, 1 \}$ & $5 \cdot 10^3$ & $10^{-6}$ & $10^{-6}$ \\
    \midrule
    &E-roads exps& 0.075 & $\{ 0.5, 0.75, 1 \}$ & $2 \cdot 10^3$  & $10^{-5}$  & $10^{-5}$ \\
    \bottomrule
    \end{tabular}
    \caption{Experimental parameters. With range($x,y,z$) we denote evenly spaced arrays that range from $x$ to $y-z$, with steps of size $z$. \label{tab:parameters}}
\end{table}

\section{Numerical Experiments}

\subsection{Non-dominated set of $\Omega$-$J$}

In Fig. 2(b) (main text) and \Cref{fig:panel_s1}(c) below we highlight the set of non-dominated points (also referred as maximal points) for $\Omega - \Omega_{0}$ against $J - J_\mathrm{OT}$.  Here, $\Omega_{0} = 0$ is a reference value corresponding to the case where there is no congestion, i.e., $\theta \to + \infty$. For our case of study, the set of non-dominated points is found as follows.

Let us introduce $ j = J - J_\mathrm{OT}$ and $\omega = \Omega - \Omega_{0}$. Each experimental run of BROT requires to fix $\theta$, $q$ and $\xi$, being the critical congestion threshold, the dropout probability, and the random noise used to initialize $w$. For a given $\theta$, we aim to find the best trade-off between $\omega$ and $j$ with respect to the parameters $q$ and $\xi$, which are responsible for the random exploration of the energy landscape of the bilevel optimization problem. Practically, this means searching the non-dominated $x_n = (\omega(q_n,\xi_n), j(q_n,\xi_n))$, where the index $n$ indicates a configuration of the parameters' initialization.

A point is $x_m$ said to be dominating $x_n$ if $x_m \neq x_n$ and $x_m \prec \ x_n$, where $x_m \prec \ x_n$ indicates that $\omega_m \leq \omega_n$ and $j_m \leq j _n$, with at least one inequality being strict. We look for and highlight those $x_n$ that are dominated by no points outputted by all experimental runs, at a fixed value of $\theta$.

\subsection{Synthetic experiments}

We show additional results that complement those of the synthetic experiments discussed in the main text. The interpretation of these results is as that of those already discussed in the manuscript, therefore, here we only present them briefly.

\begin{figure}[htpb]
\centering
\includegraphics[width=0.76\linewidth]{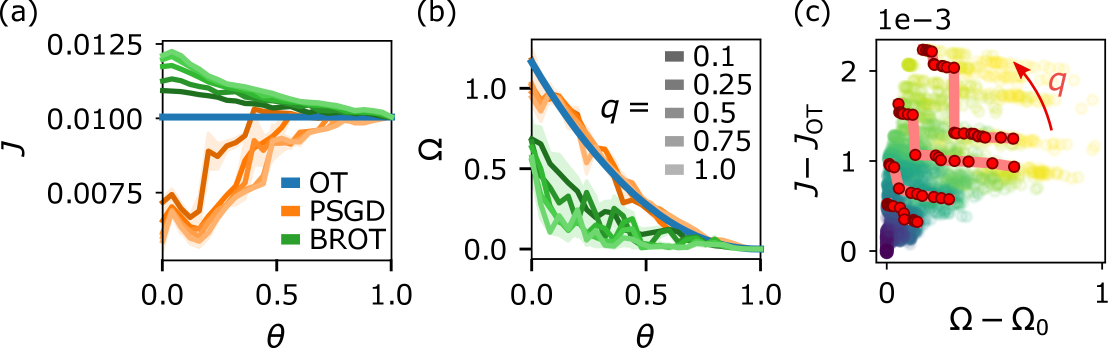}
\caption{Transport cost, congestion cost, and their trade-off for $D=4$. For an extensive caption, refer to Fig. 2 (main text).}
\label{fig:panel_s1}
\end{figure}

\begin{figure}[htpb]
\centering
\includegraphics[width=0.7\linewidth]{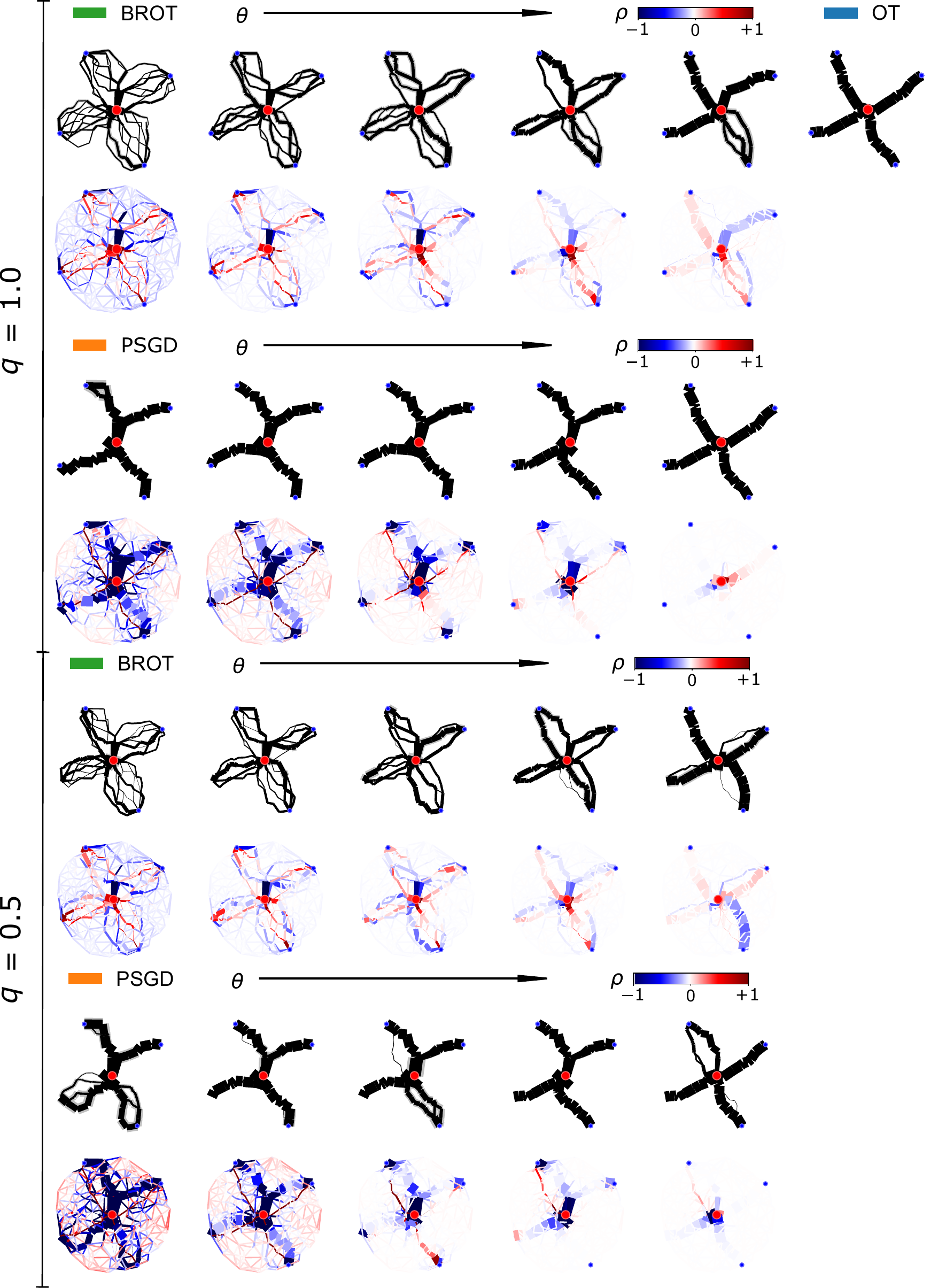}
\caption{Detailed visualization of transport networks for different methods, $D=4$ and $q=0.5,1$. For an extensive caption, refer to Fig. 2 (main text).}
\label{fig:nets4}
\end{figure}

\begin{figure}[htpb]
\centering
\includegraphics[width=0.86\linewidth]{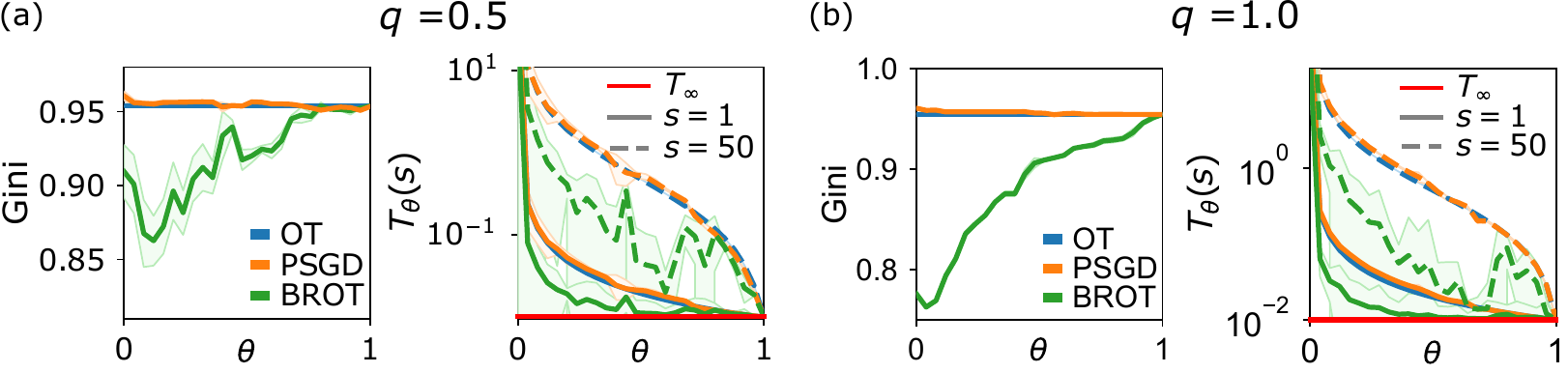}
\caption{Travel time and Gini coefficient for different methods, $D=4$ and $q=0.5,1$. For an extensive caption, refer to Fig. 3 (main text).}
\label{fig:time_gini_4}
\end{figure}

\begin{figure}[htpb]
\centering
\includegraphics[width=0.7\linewidth]{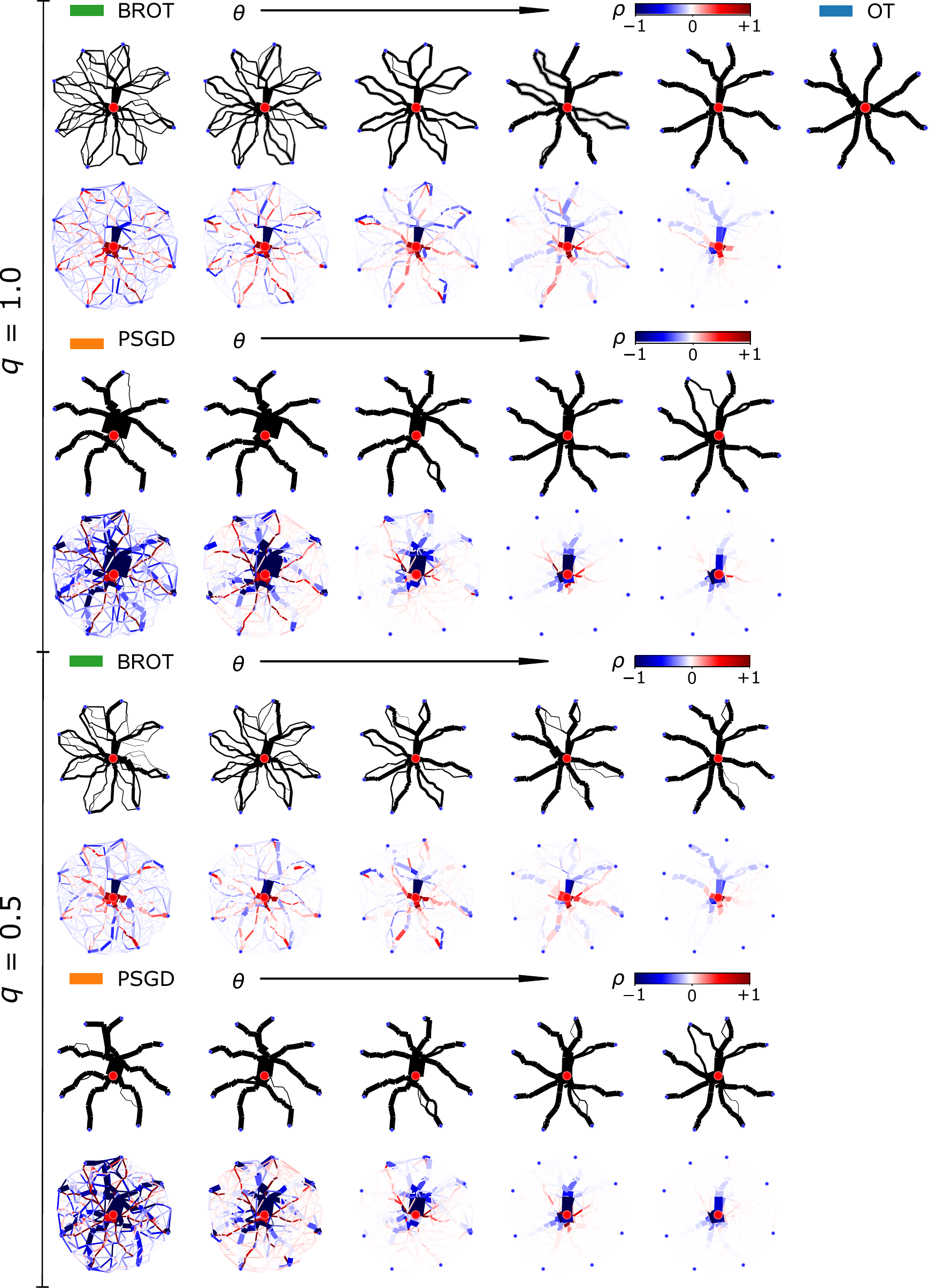}
\caption{Detailed visualization of transport networks for different methods, $D=8$ and $q=0.5,1$. For an extensive caption, refer to Fig. 2 (main text).}
\label{fig:nets8}
\end{figure}

\begin{figure}[htpb]
\centering
\includegraphics[width=0.475\linewidth]{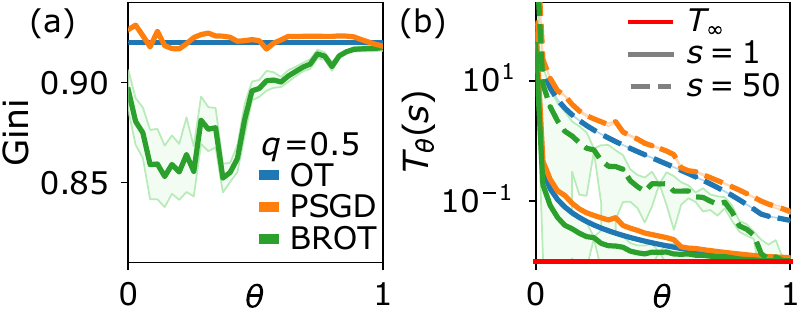}
\caption{Travel time and Gini coefficient for different methods, $D=8$ and $q=0.5$. For an extensive caption, refer to Fig. 3 (main text).}
\label{fig:time_nets8_q=0.5}
\end{figure}

First, \Cref{fig:panel_s1} and \Cref{fig:nets4} are companion figures to Fig. 2 (main text). These consist of the cost plots for $J$ and $\Omega$, of $J-J_\mathrm{OT}$ vs. $\Omega - \Omega_{0}$ for $D=4$ and $q=0.1,0.25,0.5,0.75, 1$ and of a detailed visualization of the transport networks that we extract with OT, PSGD, and BROT for $D=4$ and $q=0.5,1$. In \Cref{fig:time_gini_4} [companion Figure to Fig. 3 (main text)] we display the Gini coefficient and the total travel time $T_{\theta}(s)$ for $D=4$ and $q=0.5,1$. Similarly, \Cref{fig:nets8} and \Cref{fig:time_nets8_q=0.5} are detailed visualizations of the networks with $D=8$ and $q=0.5,1$, and a companion Figure to Fig. 3 (main text), where $q=0.5$.

It is worth mentioning how lowering $q$ accentuate the contribution of the greedy passengers, thus yielding, for the same value of $\theta$, costs that are similar to those of OT, and networks with fewer loops.

\subsection{E-road network}

\begin{figure}[htpb]
\centering
\noindent\includegraphics[width=0.5\linewidth]{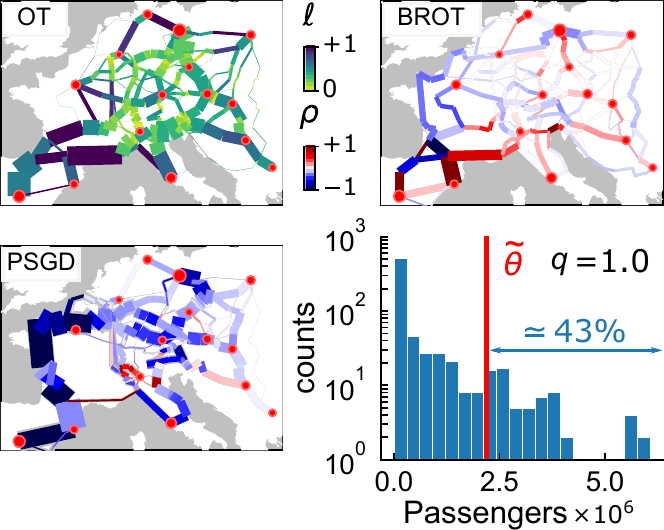} 
\caption{E-road transport networks, companion Figure to Fig. 4 (main text). Colors of edges follow the colorbars for $\ell$ (the Euclidian lengths of edges) and for $\rho$, being the difference in cost between the weights at convergence and their initial configuration.  The bottom right histogram is the shortest path distribution, computed with Dijkstra, of the fluxes. In red we mark the value of $\tilde{\theta}$ that has been used for all experiments on the E-road network, which penalizes approximately $43 \%$ of the total number of passengers traveling along their shortest path.}
\label{fig:nets_real_cost}
\end{figure}

In \Cref{fig:nets_real_cost} we show companion plots to those in Fig. 4 (main text). Particularly, we color the E-road network with the nondimensionalized Euclidean lengths $\ell$,  which are used for the initial conditions \Crefrange{eqn:label_ot_init}{eqn:label_brot_init}.  Then, we show the change of cost at convergence of the numerical schemes, i.e., $\rho_X = w^\star_{X} - \ell$, with $X =$ PSGD ($q=1$), BROT.  We also display the distribution of passengers on edges at convergence of Dijkstra's, i.e., the configuration based on which the uninformed network manager of PSGD tunes the weight. The highlighted value of $\tilde{\theta}$ is that fixed for all numerical experiments on the E-road network.

\begin{figure}[htpb]
\centering
\noindent\includegraphics[width=0.825\linewidth]{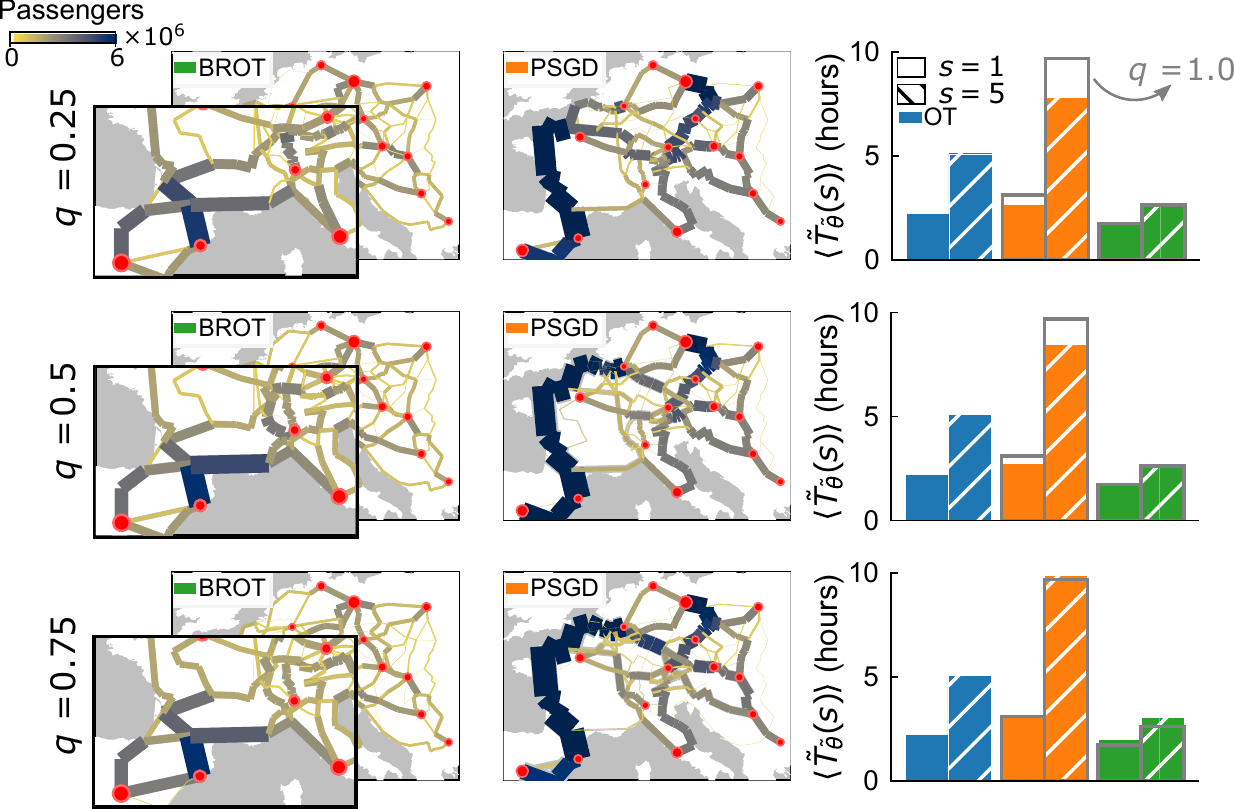} 
\caption{E-road transport networks, experiments for $q=0.25, 0.5, 0.75$. Colors of edges are the passengers travelling in the network. In the first column, we zoom into an area of interest of the networks outputted by BROT in order to highlight the presence of different network configurations that are decongested.  Colors and hatches of histograms are as in Fig. 4 (main text). In gray, we draw the average travel times for $q=1$.}
\label{fig:nets_real_q}
\end{figure}

Additionally, in \Cref{fig:nets_real_q} we display experiments on the E-road performed fixing $q=0.25,0.5,0.75$. In these we see how increasing $q$ in PSGD, the PoA (quantified by the average travel time $ \langle \tilde{T}_{\tilde{\theta}} (s) \rangle $) gets higher because of the starker lack of cooperation between greedy passengers and network manager. This yields transport networks with more congested paths. Conversely, BROT keeps the average travel time low for all $q$, i.e., it finds different local minimizers---transport networks---that are not over-trafficked.

\end{document}